\documentclass[pra,twocolumn,preprintnumbers,amsmath,amssymb]{revtex4}

\usepackage{graphicx}
\usepackage{dcolumn}
\usepackage{bm}
\usepackage{color}
\usepackage{epstopdf}
\usepackage{float}
\usepackage{ulem}

\begin{document}

\title{Temporal wave trapping from dynamical pump pulses}
\author{T. Torres$^{1,*}$, R. Terrier$^{1}$, J. Fatome$^{1}$, B. Kibler$^{1}$, G. Millot$^{1}$ and G.P. Agrawal$^{2}$}
\affiliation{$^{1}$ Universit\'e Bourgogne Europe, CNRS, Laboratoire Interdisciplinaire Carnot de Bourgogne ICB UMR 6303, 21000 Dijon, France}
\affiliation{$^{2}$ The Institute of Optics, University of Rochester, Rochester, NY 14627, United States of America}
\affiliation{$^{*}$ theo.torres@ube.fr}

\begin{abstract}
Temporal reflection in nonlinear optical fibers provides a powerful framework for manipulating light.
In this work, we theoretically and experimentally demonstrate a novel mechanism for wave trapping induced by the dynamical evolution of a single high-order soliton pulse. Experimental measurements performed in a $5$-km-long nonlinear dispersion shifted fiber confirm the coexistence of reflected, transmitted, and trapped components, in excellent agreement with theoretical predictions. These results establish a simple and versatile route toward dynamic temporal waveguiding using a single optical pulse, opening new opportunities for all-optical control and manipulation of ultrafast signals
\end{abstract}

\maketitle

\section{Introduction}
The ability to control light with light itself represents a cornerstone of modern photonics.
Among the most striking manifestations of this capability is temporal reflection, a phenomenon in which a moving refractive-index perturbation in a nonlinear medium acts as a time-dependent boundary, analogously to spatial reflection at a material interface. 
Temporal reflection in optical fibers has emerged as a powerful framework for understanding and controlling light–matter interactions in the time domain.
This effect arises in nonlinear dispersive media, where intense optical pulses induce a modification of the medium’s refractive index, creating a temporal barrier that scatters weak probe waves and enables phenomena analogous to spatial reflection, refraction, and waveguiding~\cite{plansinis2015temporal,agrawal2025propagation}.
The concept of optical horizons further extends this framework, offering a platform for exploring analogies to black hole physics and enabling novel forms of wave manipulation and frequency conversion~\cite{aguero2020hawking}. 

Temporal reflection and optical horizons, while closely related, represent distinct processes. In particular the former is a frequency conversion process requiring only a group velocity horizon (GVH) and the latter is a correlated pair-creation mechanism involving negative norm modes and both group and phase velocity horizon~\cite{jacquet2018negative,aguero2020hawking,kranas2025introduction}.
Both mechanisms have been extensively studied in the context of nonlinear photonic systems. 
GVH were first demonstrated experimentally in optical fiber in 2008~\cite{philbin2008fiber}. Following on this discovery, the temporal reflection associated with GVH was subsequently observed in fiber platforms~\cite{choudhary2012efficient,tartara2012frequency,webb2014nonlinear}, in photonic waveguides~\cite{ciret2016observation}, and in integrated circuit~\cite{khallouf2025dual}.
The negative norm modes associated with the optical black hole was observed in~\cite{drori2019observation}. In all the above experiments, it is interesting to note that ultrashort pump pulses ($<$ 1 ps) were used in order to create the GVH.

The phenomena of a temporal reflection from a single pump pulse can be naturally enriched by considering additional nonlinear effects such as Raman scattering~\cite{zhang2022temporal,agrawal2024temporal}, fined tuned dispersion control~\cite{demircan2011controlling} or temporal shaping of the pump pulse~\cite{deng2016trapping,zhang2021time,koufidis2023temporal,zhang2024spatiotemporal}. In particular, both the Raman effect 
and a pump field presenting several GVHs can lead to wave trapping due to multiple temporal reflection in the system. In particular, both form of wave trapping were observed experimentally~\cite{hill2009evolution,zhang2023experimental,zhang2023probing,wang2015bouncing} using femtoseconds pulses. 

In this work, we demonstrate a fundamentally different mechanism: the formation of trapped waves mediated by a single high-intensity pulse through its intrinsic dynamical evolution. Prior to this work, only the possibility of generating temporal reflection from a second order ultrashort soliton has been considered numerically~\cite{oreshnikov2015interaction}. Our approach bypasses the need for multiple solitons or ultrashort pulses, and opens new avenues for flexible temporal control of optical fields.

\section{Model}
\subsection{Pulse propagation}
We consider the propagation of optical pulses inside a single mode optical fiber. 
In such nonlinear dispersive medium, the propagation of the pulse's envelope can be
accurately modelled using the generalized
nonlinear 
Schrödinger
equation~\cite{agrawal2000nonlinear}:
\begin{equation}\label{eq:GNLSE}
    \frac{\partial A}{\partial z} - \sum_{m\geq2}\frac{i^{m+1}}{m!}\beta_{m}^P\frac{\partial^m A}{\partial T^m} = i\gamma |A(z,T)|^2 A(z,T),
\end{equation}
where $\gamma$ is the nonlinear parameter and $\beta_{m}^P$ are the $m$-th order dispersion at the input pulse central frequency $\omega_P$. Note that we are working in the comoving frame of the input pulse, i.e. the comoving time $T$ is related to the laboratory time $t_{\rm lab}$, via $T = t_{\rm lab} - \beta_{1}^P z$, with $\beta_{1}^P$ the inverse group velocity of the input pulse at frequency $\omega_P$. The pulse envelope $A(z,t)$ is related to the electric field as
$E(z,t) = {\rm Re}\{ A(z,t) {\rm exp}\left[i(\beta(\omega_P)z - \omega_Pt)\right]\}$.
In the following, we will be interested in pump/probe configurations where a continuous weak probe field at frequency $\omega_p$ will scatter of an intense dynamical pump field at frequency $\omega_P$. Hence, it can be convenient to separate the evolution of the probe and the pump. On the one hand the pump field, $A_P$, will obey \eqref{eq:GNLSE}. On the other hand, writing $A = A_P + ae^{i\left(\Delta\beta z - \Delta\omega T\right)}$, with $|a|\ll|A_P|$, with $\Delta \beta = \beta(\omega_p) - \beta(\omega_P)$ and $\Delta\omega = \omega_p - \omega_P$ and performing a similar expansion as for \eqref{eq:GNLSE} around the probe frequency, we obtain for the probe field
\begin{equation}\label{eq:probe_eq}
    \frac{\partial a}{\partial z} + \Delta\beta_1 \frac{\partial a}{\partial T} - \sum_{m\geq2}\frac{i^{m+1}}{m!}\beta_{m}^p\frac{\partial^m a}{\partial T^m} = 2 i\gamma |A_P(z,T)|^2 a(z,T).
\end{equation}
Note the following points in the above equation: i) the dispersion coefficients, $\beta^p_m$ are now evaluated at the probe frequency; and ii) the presence of the group velocity mismatch $\Delta\beta_1 = \beta_1(\omega_p) - \beta_1(\omega_P)$ due to the choice of comoving frame. The probe frequency satisfying $\Delta\beta_1$ defines the group velocity matched (GVM) frequency. 
%
\subsection{Eikonal approximation}
To gain further insight into the temporal reflection process, we now introduce the eikonal approximation~\cite{synge1963hamiltonian}. We look for solution of \eqref{eq:probe_eq} with a rapidly oscillating phase, i.e. we write $a(z,t) = {\rm exp}\left[iS(z,T)/\epsilon\right]$~\footnote{Here $\epsilon$ is just an order counting parameter.}. We further assume that all variations occur on the scale much larger than the wavelength, which can be expressed by rescaling the derivative $\partial \rightarrow \epsilon \partial$. Performing a perturbative expansion, the leading order term leads to
\begin{equation}\label{eq:H-J}
    \frac{\partial S}{\partial z} + \Delta\beta_1 \frac{\partial S}{\partial T} - \sum_{m\geq2} \frac{(-1)^m}{m!} \beta_m^p \left( \frac{\partial S}{\partial T}\right)^m = 2\gamma |A_P(z,T)|^2.
\end{equation}
This nonlinear first order PDE can be solved using the method of characteristics, which are the \textit{rays} of the wave. To obtain the trajectories of the rays, we solve Hamilton's equation after substituting $\partial_z S = k_z$ and $\partial_T S = \Omega$ into \eqref{eq:H-J} which gives the Hamiltonian of the system,
\begin{eqnarray}
    \mathcal{H}(z,T,k_z,\Omega) = k_z + &\Delta&\beta_1\Omega - \sum_{m\geq2} \frac{(-1)^m}{m!}\beta_m^p \Omega^m \nonumber \\
    &-& 2\gamma|A_P(z,T)|^2. \label{eq:Ham}
\end{eqnarray}
One of Hamilton's equations directly expresses the cross phase modulation between the pump and the probe\footnote{Note that usually, the rays are parametrized by an independent parameter, say $\sigma$. However, here we can choose to parametrize the rays by $z$ directly, because $\mathcal{H}$ is linear in $k_z$. This linear relation implies that $\frac{dz}{d\sigma} = 1$, hence $z$ can be chosen as an affine parameter.}
\begin{equation}\label{eq:XPM}
\frac{d\Omega}{dz} = -\frac{\partial\mathcal{H}}{\partial T} = 2\gamma \partial_T|A_P(z,T)|^2.
\end{equation}
We see that temporal reflections will be possible provided that the gradient of the intensity is sufficiently strong. That is why experiments related to temporal reflection and optical horizons have mainly used ultrashort pulse ($<1$ ps). However, from \eqref{eq:XPM}, we see that it is the instantaneous gradient that matters and that the pump field does not necessarily need to exhibit a strong gradient during the entire propagation. This observation opens the door to using pump fields whose properties evolve during the propagation.
We explore this avenue in the following by considering pump pulses with a dynamical evolution similar to the one of higher-order solitons. In particular we focus on few picoseconds long pulses with a dynamical evolution capable of reproducing temporal reflection from ultrashort pulses as well as wave trapping from solitonic cages.

\section{Temporal reflection and wave trapping}
\subsection{Theoretical description}

We begin our discussion of the temporal reflection and wave trapping phenomena by numerically studying the dynamics of a continuous weak probe field scattering of a dynamical pump field. 
We evolve \eqref{eq:GNLSE} with the initial condition $A(z=0,t) = A_P(t) + a e^{i\Delta\omega t}$, where $A_P(t) = \sqrt{P_0}\rm{sech}(t/T_s)$ and $0<a\ll1$. 
Anticipating the experimental results of the next section, we work in a regime similar to the one accessible experimentally and we take $T_s = 3.77$ \rm{ps} and $a = 0.5$ mW. 
The nonlinear coefficient $\gamma$ is taken to be $1.8\ \rm{W}^{-1}.\rm{km}^{-1}$ and we limit ourselves to the dispersion coefficients $\beta_2^P \approx -1.52 \ \rm{ps^2.km^{-1}}$, $\beta_3^P \approx 0.12\ \rm{ps^3.km^{-1}} $ and $\beta_4^P = -0.0006\ \rm{ps^4.km^{-1}}$ at the pump wavelength \mbox{$\lambda_P = 1560\ {\rm nm}$}. 
These parameters correspond to the dispersion shifted fiber (DSF) that will be used for the experimentally observation of the predicted effect. 
We evolve our field over a distance $z=5\ \rm{km}$ which corresponds to the length of DSF at our disposal.
We work in the reference frame of the pump field and the frequencies are expressed relative to the pump carrier frequency. With these parameters, higher order solitons will have a periodic evolution with period $z_0 = \frac{\pi}{2}\frac{T_s^2}{|\beta_2^P|} \approx 14.7\ \rm{km}$. 
\begin{figure}[h!]
    \centering
    \includegraphics[width=0.9\linewidth]{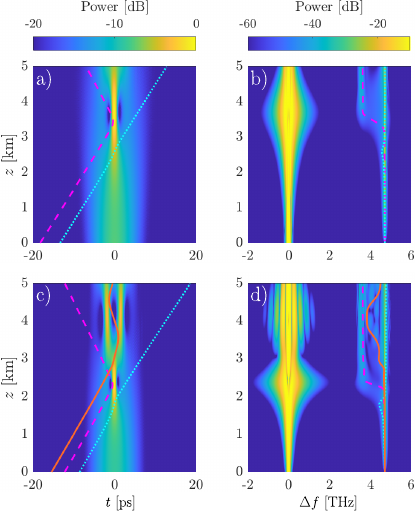}
    \caption{\textbf{Numerical demonstration of temporal reflection and wave trapping from a dynamical pump.} Panel a) and b) show the temporal evolution of the pump pulse and the spectral evolution along the fiber respectively for an input pump power $P_0 \approx 600\ {\rm mW}$. Two rays with different delays are plotted in panel a): i) a reflected ray (dashed purple) that hits the pump at the maximum compression and ii) a transmitted ray (dotted blue) that reach the pump before the compression.
    Panel c) and d) are the equivalent to panel a) and b) with an input pump power $P_0 \approx 850\ {\rm mW}$. Three rays with different delays are plotted in panel a): i) a reflected ray (dashed purple); ii) a transmitted ray (dotted blue) and iii) a trapped ray (solid red) that undergoes multiple reflection in the bipulse.}
    \label{fig:ray}
\end{figure}
Working with a second order soliton, will lead to a maximum compression occurring at $z=z_0/2$ which is larger than the fiber length available. 
Hence we will work with input power corresponding to third and fourth order solitons. 
In particular we will aim at having pump fields which exhibit either a single compression stage (Fig.~\ref{fig:ray}.a ) or a compression stage followed with the formation of bi-pulse (Fig.~\ref{fig:ray}.c ), over the length of the fiber. Since we are working with a continuous probe field, we are mainly interested in the spectral evolution during the propagation.
Fig.~\ref{fig:ray} shows the temporal evolution of the pump field (panels a and c) as well the spectral evolution of both the continuous probe (with $\Delta f = \Delta \omega/(2\pi) \approx 4.6\ {\rm THz}$ and pump field (Panels b and d). The top (bottom) panels correspond to an input power, $P_0 \approx 600 \ {\rm mW}$ ($850\ {\rm mW}$). 
\\
In the case $P_0 \approx 600 \ {\rm mW}$, we see that the pump field undergoes a single temporal compression stage over the length of the fiber, occurring around $z\approx3750\ {\rm m}$. 
With this temporal compression is associated a spectral broadening (shown in Fig.~\ref{fig:ray}.b) ) and the generation of a red shifted wave corresponding to the temporal reflection of the probe field. 
Hence the temporal reflection mainly occurs during the temporal compression of the pump field. 
This is highlighted by computing two different rays from the Hamiltonian \eqref{eq:Ham} with two different initial delays. 
One ray is chosen with a delay such that it will hit the pump field at this maximum compression stage while the other is launched with a smaller delay and will reach the pump pulse before the compression stage. We can clearly see, both temporally and spectrally, that it is the former ray that gets temporally reflected while the latter is transmitted.
\\
In the case $P_0 \approx 850 \ {\rm mW}$, the pump field undergoes a temporal compression ($z\approx 2400\ {\rm m} )$ followed by the formation of a bipulse ($z\approx 4000 \ {\rm m}$) over the length of the fiber. %
The temporal compression is associated to spectral broadening and the bipulse formation to modulated spectrum (shown in Fig.~\ref{fig:ray}.d) ).
As in the previous case, when the pump field is compressed, a temporally reflected wave is generated.
In addition, with the formation of the bipulse, new frequencies are generated around the group velocity matched frequency, i.e. $\Delta\beta_1(\omega) = 0$. These frequencies are associated with parts of the probe field that gets trapped between the bipulse, a situation similar to the soliton cage.
This is again highlighted by computing different rays with three different initial delays. 
%
\subsection{Experimental setup}\label{sec:Exp}
\begin{figure}[!h]
    \centering
    \includegraphics[width=0.9\linewidth]{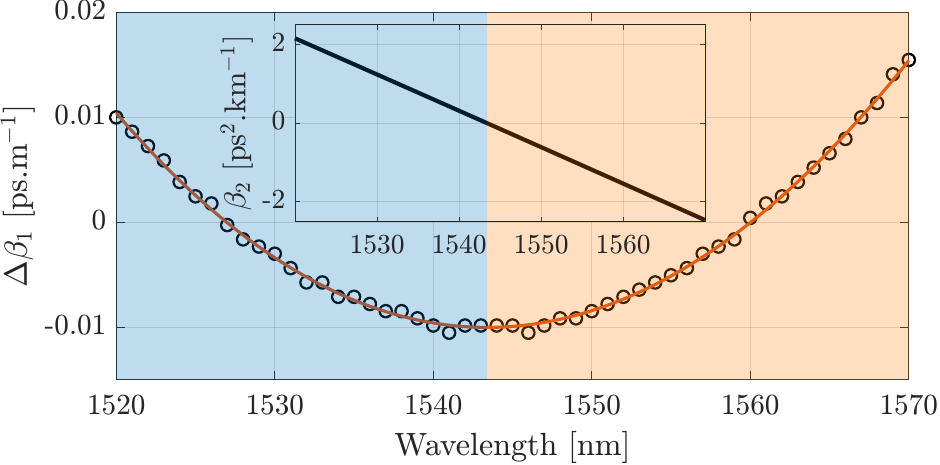}
    \caption{\textbf{Dispersion measurement of the dispersion shifted fiber.} Experimental measurement of the inverse relative group velocity $\Delta\beta_1$ (round markers) and its fit with a third order polynomial in frequency (red line). At the pump wavelength, $\lambda_P = 1560\ {\rm nm}$  we obtain the following dispersion parameters : $\beta_2^P \approx -1.52 \ \rm{ps^2.km^{-1}}$, $\beta_3^P \approx 0.12\ \rm{ps^3.km^{-1}} $ and $\beta_4^P \approx -0.0006\ \rm{ps^4.km^{-1}}$; those are used in the simulation throughout the study. The inset represents the second order dispersion parameter $\beta_2$ as a function of the wavelength, estimated from the fit.}
    \label{fig:dispersion}
\end{figure}

We now verify the above prediction by comparing the results of our numerical simulation with experimental results.
The setup is composed of a mode-locked fiber laser delivering pulse with nearly ${\rm sech}^2$ shape at a 40 MHz repetition rate and a temporal full width at half maximum around $6.7\ {\rm ps}$ at a carrier wavelength of $1560\ {\rm nm}$. 
An optical attenuator is placed after the source in order to control the ''order" of the pump solitonic pulse. 
The pump field is combined with a continuous wave probe with a tunable wavelength ranging from $1520\ {\rm nm}$ to $1535\ {\rm nm}$, before being injected into a $5$-km-long DSF. 
The relative inverse group velocity $\Delta\beta_1$ is measured experimentally and from it, we extract the higher order dispersion coefficients.
Fig.~\ref{fig:dispersion} shows $\Delta\beta_1$ and $\beta_2$ as functions of the wavelength. 
%
%
%
\subsection{Emergence of temporal reflection and wave trapping}
\begin{figure}[!h]
    \centering
    \includegraphics[width=\linewidth]{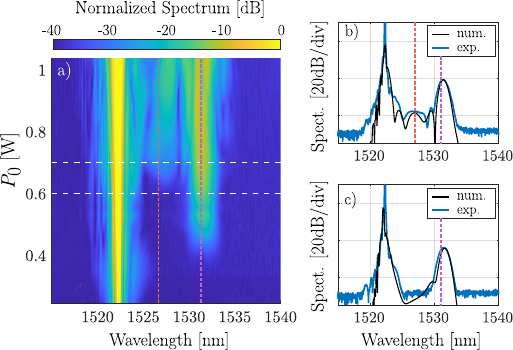}
    \caption{\textbf{Emergence of temporal reflection and wave trapping.} 
Panel a) shows the optical spectrum at the output of the DSF for different pump peak powers for a continuous probe centred around $1522\ {\rm nm}$. 
The pump pulse is centred around $1560\ {\rm nm}$. Panels b) and c) show the measured (blue) and simulated (black) spectra at the fiber end for $P_0 = 700 \ {\rm mW}$ and $600\ {\rm mW}$ respectively.
The vertical purple and red dashed lines show the wavelength of the temporal reflection and the group velocity matched wave predicted from the dispersion measurements shown in Fig.\ref{fig:dispersion}. 
}
    \label{fig:power_scan}
\end{figure}

We first observe the influence of the input pump peak power on the output spectrum at a fixed probe wavelength around $1522\ {\rm nm}$. Fig.~\ref{fig:power_scan} a) shows the optical spectrum at the output of the DSF as the pump power is varied from $250 \ {\rm mW}$ to $1.04\ {\rm W}$. 
At low enough power, \mbox{$P_0 \lesssim 400\ {\rm mW}$}, only the input probe wavelength is present, i.e. no frequency conversion occurs during the propagation. 
As the power increases, for $400\ {\rm mW}\lesssim P_0 \lesssim 600\ {\rm mW}$, we  see the emergence of a blue shifted peak around $1531\ {\rm nm}$, corresponding to the wavelength of the temporally reflecting pulse.
Then, for larger power, $P_0 \gtrsim 600 \ {\rm mW}$, we observe the apparition of third peak at a the group velocity matched wavelength, corresponding to the predicted trapped wave. 
Figs.~\ref{fig:power_scan}.b) and c) show the measured output optical spectrum compared with the results of the numerical simulations of \eqref{eq:GNLSE} for input pump power of $600 \ {\rm mW}$ and $700 \ {\rm mW}$, respectively. Both measurements and simulations shows the distinct peaks associated with the temporal reflection and the wave trapping effect.

\subsection{Wavelength dependency}

\begin{figure}[h!]
    \centering
    \includegraphics[width=1\linewidth]{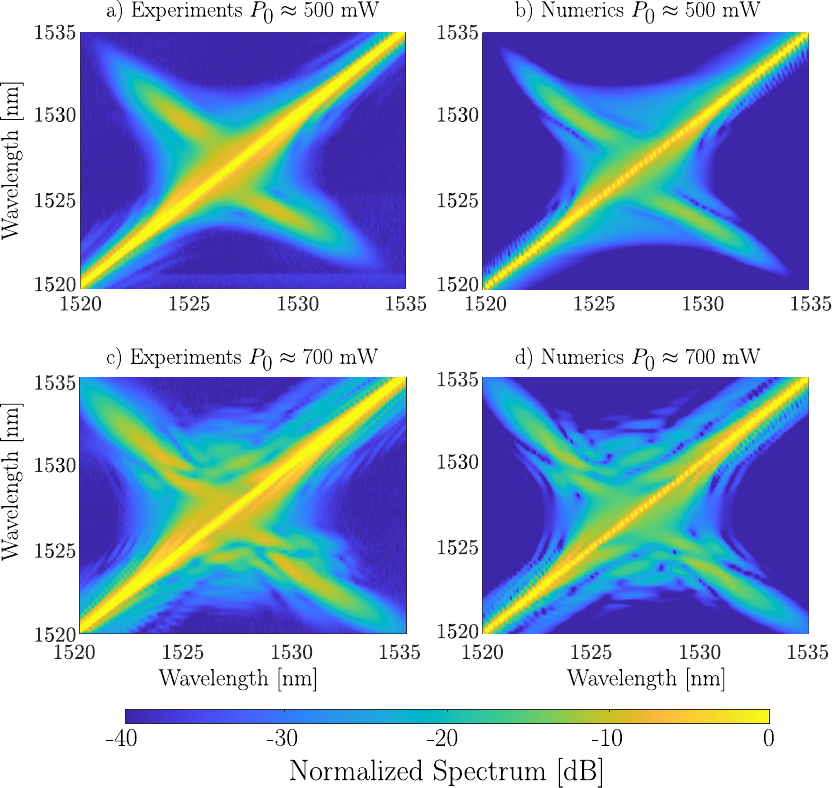}
    \caption{\textbf{Comparison between frequency conversion processes for two distinct dynamical evolution of the pump field.} The probe input wavelength varies from $1520\ {\rm nm}$ to $1535\ {\rm nm}$. The top (bottom) panels show the experimental and numerical output spectra for an input pump power $P_0\approx 500\ {\rm mW}$ ($P_0\approx 700\ {\rm mW}$). Note the asymmetry of the ''cross-like" pattern as well as the presence of additional substructure in the experimental spectrum which is qualitatively reproduced by the numerical simulations.}
    \label{fig:X_plot}
\end{figure}

We now turn our attention to the wavelength dependency of the processes. 
Again we consider scenarios where the pump field undergoes either a single compression or a compression followed by a bipulse. We chose the input power for these two scenarios to be $P_0 \approx 500 \ {\rm mW}$ and $P_0 \approx 700 \ {\rm mW}$ respectively.
We then vary the probe wavelength from $1520 \ {\rm nm}$ to $1535 \ {\rm nm}$.
Figs.~\ref{fig:X_plot}~a) and b) show the experimental and numerical output spectra for $P_0 \approx 500 \ {\rm mW}$. We observe the typical ''cross-like" pattern: The diagonal branches correspond to temporal reflection, while the centre of the cross corresponds to the group velocity matched wavelength of the trapped wave. We note that these maps present two interesting features: i) the ''cross" is asymmetrical close to the group velocity matched wavelength, and ii) the presence of a substructure, in particular spectral dips (or other modulation) in the reflected branches. We intuitively attribute the former to the finite propagation distance 
that implies that the probe still undergoes the XPM from the pump pulse.
The latter effect might reflect the internal dynamics of the pump pulse.
\\
Figs.~\ref{fig:X_plot}~c) and d) show the experimental and numerical output spectra for $P_0 \approx 700 \ {\rm mW}$.
We recover the ''cross-like" pattern of Panels a) and b) but this time with a stronger and more intricate substructure certainly related to higher order soliton number of the pump.
Understanding the detailed characteristics of the ''cross-like" pattern is warranted and left for future work.

\section{Conclusion}

In this work, we have theoretically and experimentally demonstrated a novel mechanism for temporal wave trapping in nonlinear optical fibers, using the intrinsic dynamical evolution of a single high-intensity pump pulse. 
By exploiting the natural compression and temporal reshaping of higher-order solitons, we overcome the need for ultrashort pulses ($<$ 1 ps) or complex multi-soliton configurations previously  used for generating temporal reflection and wave trapping. 
Our results reveal that a single, few picoseconds long pump pulse can induce both temporal reflection and group-velocity-matched trapping of a continuous weak probe wave, with the trapped component arising from multiple reflections within a dynamically formed bipulse.
This interpretation is further supported by the ray tracing analysis which allows for a deeper physical understanding of these mechanisms.

These results pave the way for dynamic, reconfigurable temporal waveguiding in nonlinear media, with potential applications spanning from telecommunications to quantum optics.

\medskip
\noindent
\textbf{Acknowledgment} The authors thank E. Lucas for valuable discussions and insights during the experimental phase.
\\
\textbf{Disclosures} The authors declare no conflicts of interest.
\\
\textbf{Data availability} Data underlying the results presented in this paper may be obtained from the authors upon reasonable request.

\bibliography{sample}


\end{document}